\documentclass{revtex4}
\usepackage{graphicx}
\usepackage{amsmath}
\usepackage{amssymb}
\usepackage{yfonts}
\usepackage{braket}

\DeclareMathOperator{\erf}{erf}

\begin{document}

\title{Quantum optical waveform conversion}

\author{D. Kielpinski$^{1,3}$, J. F. Corney$^4$, and H. M. Wiseman$^{2,3}$}
\affiliation{$^1$ ARC Centre of Excellence for Coherent X-Ray Science\\
$^2$ ARC Centre of Excellence for Quantum Computer Technology\\
$^3$ Centre for Quantum Dynamics, Griffith University, Nathan QLD 4111, Australia\\
$^4$ ARC Centre of Excellence for Quantum-Atom Optics, School of Mathematics and Physics, University of Queensland, St.~Lucia QLD 4072, Australia}

\maketitle
\setlength{\parindent}{0in}

\textbf{Currently proposed architectures for long-distance quantum communication rely on networks of quantum processors connected by optical communications channels \cite{Briegel-Zoller-quantum-repeater, Duan-Zoller-linear-optics-atom-QC}. The key resource for such networks is the entanglement of matter-based quantum systems with quantum optical fields for information transmission. The optical interaction bandwidth of these material systems is a tiny fraction of that available for optical communication, and the temporal shape of the quantum optical output pulse is often poorly suited for long-distance transmission. Here we demonstrate that nonlinear mixing of a quantum light pulse with a spectrally tailored classical field can compress the quantum pulse by more than a factor of 100 and flexibly reshape its temporal waveform, while preserving all quantum properties, including entanglement. Waveform conversion can be used with heralded arrays of quantum light emitters to enable quantum communication at the full data rate of optical telecommunications.} \\

The development of long-distance quantum communication is critical for future quantum cryptography and distributed quantum computing applications. Current fiber-optical quantum communication systems rely on direct transmission of quantum light pulses, but the attenuation of the fiber imposes a distance limit of tens of kilometers for this kind of quantum communication \cite{Gisin-Zbinden-quantum-crypto-rev} by virtue of the no-cloning theorem \cite{Wootters-Zurek-no-cloning}. Quantum repeater architectures \cite{Briegel-Zoller-quantum-repeater, Duan-Zoller-linear-optics-atom-QC} promise to circumvent this limit by preparing entangled states over an optical communications channel and storing these entangled states as a resource for subsequent quantum communication. Components of quantum repeaters have now been demonstrated with a wide variety of physical systems, including single atoms \cite{Blinov-Monroe-ion-photon-entanglement, McKeever-Kimble-CQED-single-photon, Keller-Walther-ion-cavity-single-photon}, atomic vapors \cite{Chou-Kimble-atomic-ensemble-single-photon, Julsgaard-Polzik-atom-quantum-memory-light, Hosseini-Buchler-coherent-optical-pulse-sequencer}, rare-earth ions in solids \cite{deRiedmatten-Gisin-rare-earth-photon-storage, Hedges-Sellars-efficient-quantum-memory}, quantum dots \cite{Yilmaz-Imamoglu-quantum-dot-photon-entanglement}, and NV-centres \cite{Togan-Lukin-NV-photon-entanglement}. The common thread among these demonstrations is the manipulation of quantum light pulses by matter-based quantum emitters. The temporal waveform of such emitters is typically a single-sided exponential with decay constant on the order of 1 nanosecond, which cannot be readily mode-matched to the smooth, broadband pulses desirable for telecommunications, posing a substantial disadvantage for quantum-emitter approaches to quantum networking.  Recent attempts to overcome these issues include increasing the emitter bandwidth to the GHz range by the use of nonresonant interactions \cite{Reim-Walmsley-fast-atom-photon-memory},  shaping the temporal waveform by placing the emitter in a nonlinear resonator \cite{McCutcheon-Loncar-quantum-pulse-cavity-shaping}, temporal modulation of single-photon wavepackets \cite{Specht-Rempe-single-photon-phase-shaping, Belthangady-Harris-single-photon-spectral-hiding}, and nonlinear frequency conversion experiments with single photons \cite{Vandevender-Kwiat-single-photon-upconversion, Albota-Wong-telecom-upconversion-detection, Langrock-Fejer-telecom-upconversion-detection, Tanzilli-Zbinden-entanglement-preserving-upconversion}. \\

\begin{figure}
\includegraphics*[width=15cm]{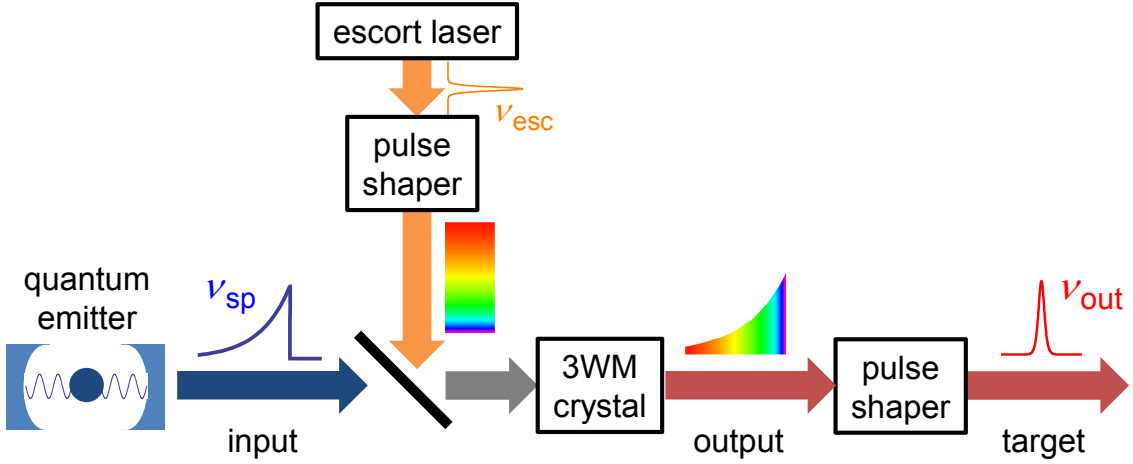}
\caption{Schematic of the quantum optical waveform converter. A nonclassical light input (originating, e.g., from a quantum emitter in a high-finesse optical resonator) is combined with a highly chirped classical pulse. The combined fields undergo three-wave mixing (3WM) in a nonlinear crystal, transferring the spectral modulation of the classical pulse onto the 3WM output. The output is separated from the original fields and passed through a pulse shaper to remove the residual phase, producing a pulse in the target mode that inherits the quantum state of the input mode. The colours under the pulse envelope represent the frequency variation during the pulse length, with the variation of colours greatly exaggerated for clarity.}
\label{schematic}
\end{figure}

Here we present an efficient and straightforward method of quantum optical pulse shaping and compression that immensely simplifies the interfacing of quantum emitters with telecommunications networks. Our method preserves the full quantum statistics of the input field, including entanglement and any other multimode correlations, while enabling compression by more than a factor of 100, along with flexible reshaping of the temporal waveform. In particular, our method enables time/wavelength transduction of spontaneously emitted photons from quantum emitters into short, smooth pulses at telecommunications wavelengths. As shown in Fig. \ref{schematic}, the input field undergoes three-wave mixing (3WM) with a frequency-chirped classical laser pulse. For an appropriate choice of classical laser intensity and chirp, the 3WM product radiation has the same spectrum as the desired target mode, but receives the quantum statistics of the input. The 3WM output is then dechirped with a second pulse shaper to match the temporal wavefunction of the target mode. The waveform converter extends the classical time-lens technique, which has achieved remarkable results in compressing and stretching classical pulses \cite{Bennett-Kolner-3WM-time-lens-100x, Foster-Gaeta-time-lens-waveform-compression}, to the quantum domain and to arbitrary pulse reshaping. \\

\begin{figure}
\includegraphics*[width=15cm]{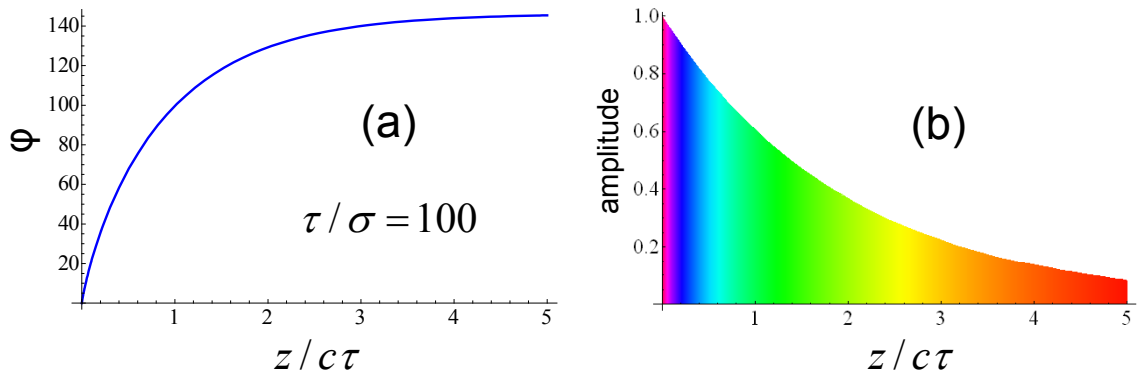}
\caption{a) Escort phase modulation function $\phi(z)$ for conversion from a single-sided exponential waveform to a Gaussian waveform with compression ratio $\tau/\sigma = 100$. b) Visualisation of the spectral chirp $d\phi(z)/dz$ imposed on the initial waveform by 3WM. Height indicates waveform amplitude, colour indicates local frequency after escort phase imprinting. The hue of the colour is proportional to $d\phi(z)/dz$ as calculated from Eq. (\ref{Gaustarg}).}
\label{chirpfig}
\end{figure}

We describe the 3WM process using slowly varying bosonic field operators $\Psi_j(z,t)$ with $j=1$ the input mode and $j=2$ the mode generated by 3WM. Here $z$ measures the distance along the propagation axis in a frame comoving at the group velocity, and $t$ measures the duration of the interaction between the three fields. (This coordinate convention is to be contrasted with the classical nonlinear optics convention in which $z$ measures the interaction length in the 3WM medium and $t$ is the time of arrival at the detector.) The escort laser pulse contains $\gtrsim 10^{10}$ photons and can be approximated as a classical field that remains unaffected by 3WM. For an escort pulse much longer than the other pulses, the Hamiltonian becomes (see Methods)
\begin{equation}
H_\mathrm{3WM} = i \hbar \Omega \int dz \: e^{i \phi(z)} \Psi_1^\dagger \Psi_2 + \mbox{h.c.} \label{hamid}
\end{equation}
where $\Omega$ is determined by the nonlinear coupling constant and the intensity of the escort pulse, and where $\phi(z)$ is the phase of the escort field. The quantum field operators evolve as
\begin{align}
\Psi_1(z,t) &= \cos \Omega t \Psi_1(z,0) + \sin \Omega t \: e^{i \phi(z)} \Psi_2(z,0) \label{qfideal1} \\
\Psi_2(z,t) &= -\sin \Omega t \: e^{-i \phi(z)} \Psi_1(z,0) + \cos \Omega t \Psi_2(z,0) \label{qfideal2}
\end{align}
If the fields leave the 3WM medium after an interaction time $T = \pi/(2 \Omega)$, the solution for mode 2 at times $t > T$ is just $\Psi_2(z,T) = - e^{-i\phi(z)} \Psi_1(z,0)$. The quantum state of mode 1 is perfectly transferred into mode 2, while mode 2 has acquired the phase $-\phi(z) + \pi$ from mode 3. To temporally match the 3WM output to the target pulse shape, the output pulse shaper then removes the undesired relative phases of the spectral components, performing a unitary transformation on the output field operator. The spontaneous emission from a quantum emitter, with a single-sided exponential waveform, can be converted into a much shorter Gaussian pulse by choosing (see Methods)
\begin{align} \label{Gaustarg}
\phi(z) = \frac{\sqrt{2}}{\sigma} \int_0^z d\zeta \: \erf^{-1}(e^{-\zeta/(c \tau)})
\end{align}
where $\tau$ is the spontaneous emission lifetime and the target amplitude is proportional to $e^{-z^2/(2 (c \sigma)^2)}$. The phase modulation of Eq. (\ref{Gaustarg}) is visualised in Fig. \ref{chirpfig}. \\

The ideal quantum waveform conversion described above will not be achieved with unit fidelity in real 3WM media because of dispersion. We
now show that the fidelity $F$ can nevertheless exceed 99.9\% for readily achievable experimental parameters. For pure input states, $F = \left |\Braket{\psi_\mathrm{ideal} | \psi_\mathrm{disp}}\right|$, where the result of ideal evolution is written $\ket{\psi_\mathrm{ideal}}$ and the result with dispersion included is $\ket{\psi_\mathrm{disp}}$. If the input system is entangled with another quantum system, the fidelity of the final entangled state is simply the average fidelity of the eigenstates of the input density operator, weighted by their corresponding eigenvalues. A perturbative analysis of the dispersive evolution (see Methods) shows that the error is dominated by mismatch between the group velocities $v_{1,2,3}$ of the input, output, and escort fields, which we parametrise by $v = (v_1-v_2)/2$ and $v_e = v_3 - (v_1+v_2)/2$. For a pure input wavefunction $A(z)$ with characteristic length scale $L$, we define dimensionless velocities $u = v/v_0$, $u_e = v_e/v_0$, where $v_0 = \Omega L/(2 \pi)$. The error can be minimised by adding the compensation phase
\begin{equation}
\Delta_\mathrm{opt}(z) = \frac{1}{8} (u_e - u) \phi^\prime(z) L \label{phscomp}
\end{equation}
to the initial escort phase $\phi(z)$. For an average photon occupation $\langle n \rangle$, we then obtain
\begin{equation}
1 - F_\mathrm{opt} = \frac{\langle n \rangle u^2 L^2}{32 \pi^2} \int dz \: \big| 2 A^\prime(z) - i (1 + u_e/u) \phi^\prime(z) A(z) \big|^2 \label{fidelity}
\end{equation}
The ratio $u_e/u$ is set by the crystal dispersion alone, but $u$ varies with the escort laser intensity $I_\mathrm{esc}$ as $u \propto I_\mathrm{esc}^{-1/2}$. Thus the fidelity can be made arbitrarily close to 1 by increasing the escort laser power. Eq. (\ref{fidelity}) can then be rewritten as $1 - F_\mathrm{opt} = (u/u_\mathrm{err})^2$ for some $u_\mathrm{err} \lesssim 1$. It can be seen from Eq. (\ref{fidelity}) that the perturbation theory breaks down for a pulse with an arbitrarily sharp leading edge; a perturbative analysis in momentum space shows that $1 - F_\mathrm{opt} \propto u$ in this limit. In practice, the time required to excite a quantum emitter is never exactly zero, so the leading edge of the pulse is smoothed over the excitation timescale. We also perform a full numerical simulation for a single-mode single-photon input state (see Methods). This confirms that the effect of group-velocity dispersion is insignificant in all cases of interest. \\

\begin{figure}
\includegraphics*[width=15cm]{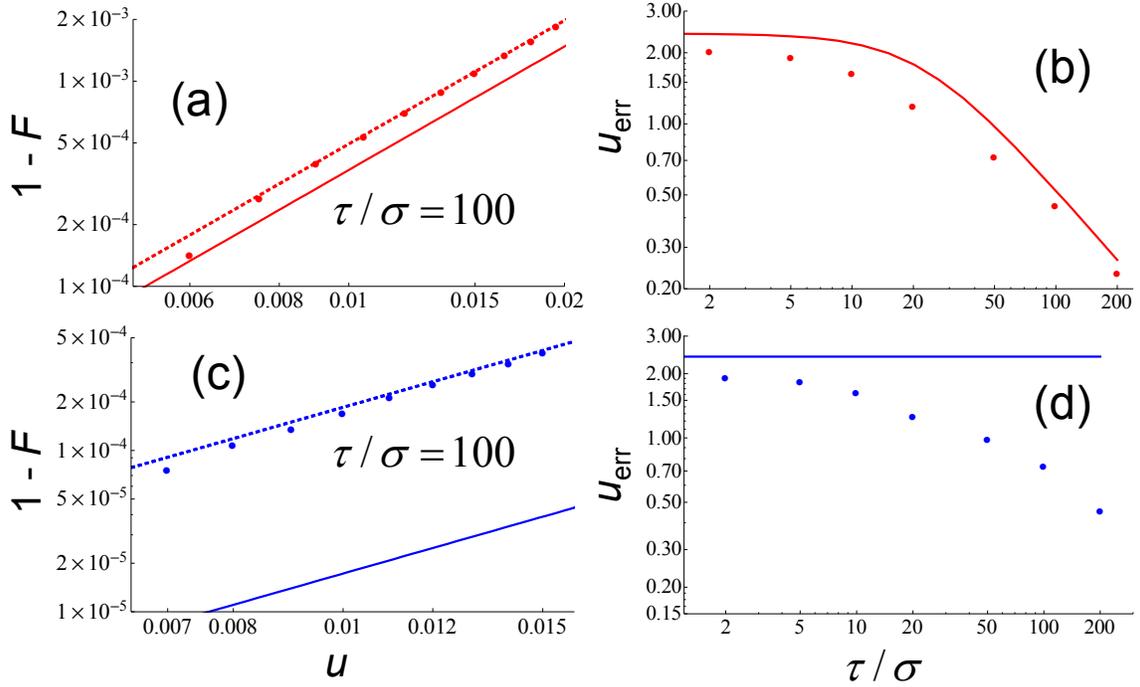}
\caption{Error induced by quantum waveform conversion from a single-sided exponential pulse of time constant $\tau$ and rise time $0.02 \tau$ to a Gaussian pulse of $1/e^2$ time constant $\sigma$. Errors are less than $10^{-3}$ for readily achievable experimental parameters (see text). a) Error at compression ratio $\tau/\sigma = 100$ as a function of dimensionless group-velocity mismatch $u$ for conversion of 370 nm photons to 1550 nm in lithium niobate. Solid line: perturbative prediction. Points: simulation results. Dashed line: Best fit of $1-F = (u/u_\mathrm{err})^2$ to simulation points. b) Error scale $u_\mathrm{err}$ as a function of compression ratio $\tau/\sigma$. Solid line: perturbative prediction. Points: simulated values computed from least-squares fits to simulation results. As expected, the simulation results match well to the perturbative theory. c), d) are the same as a), b), but for conversion of 780 nm photons to 1550 nm in type-II matched lithium niobate, near the special point $v = -v_e$ at which the escort phase term in Eq. (\ref{fidelity}) vanishes. Errors are even lower than for conversion of 370 nm photons, but at compression ratio above 10 the perturbation theory breaks down and higher-order GVM dominates the error.}
\label{errfig}
\end{figure}

Figure \ref{errfig} shows the analytic and numerical error estimates for two cases of particular experimental interest. Case 1: the conversion of 370 nm photons from a $\mbox{Yb}^+$ ion \cite{Blinov-Monroe-ion-photon-entanglement} to the 1550 nm telecommunications band using periodically poled lithium niobate, for which $u_e/u \approx -2/3$. The simulated error closely follows the perturbative result for small values of $u$ up to a compression ratio $\tau/\sigma = 200$. For an escort laser pulse of energy $\sim 1 \mu$J and duration of 150 ns ($> 20 \: \tau_\mathrm{Yb^+}$) in a 50 mm long crystal waveguide, one finds $u = 0.013$ and error of $1-F = 7 \times 10^{-4}$ at compression ratio of 100. Case 2: the conversion of 780 nm photons from a Rb atom to the telecommunications band, for which one can arrange $u_e/u = -1$ by poling lithium niobate for type II phase matching \cite{Yu-Taira-PPMgLN-GVM-matching}. Here the error is much lower for similar escort laser parameters. $v < 10^5 \mbox{m}/\mbox{s}$ for any choice of output wavelengths in the telecommunications band, so $u < 10^{-4}$ and $1-F \ll 10^{-4}$.  \\

We have shown that three-wave mixing with a modulated classical field can reshape and compress the waveform of a quantum light pulse while faithfully maintaining the quantum information carried by the photons. A quantum light pulse produced by a quantum emitter with a lifetime of nanoseconds can be converted to a Gaussian pulse with a duration of tens of picoseconds that is compatible with standard telecommunications protocols. The low error of waveform conversion is compatible with schemes for fault-tolerant quantum communication over long distances \cite{Briegel-Zoller-quantum-repeater}. Quantum waveform conversion enables simultaneous time- and wavelength-division multiplexing of the pulses from an array of quantum emitters up to the limit of channel capacity, massively increasing quantum communications bandwidth. Current DWDM systems with 50 GHz channel spacing achieve their maximum capacity for transform-limited pulses with $\sim 20$ ps duration, while the dispersive effects of long-haul fibre transmission require the pulses to have a smooth temporal waveform. As each pulse arrives from the emitter array, it can be simultaneously converted to this ideal waveform and sorted into an appropriate DWDM channels. The rate of entangled pair generation in a quantum network is then limited only by the telecommunications bandwidth and the size of the emitter array. \\

\noindent \textbf{Methods}\\
\begin{small}

\noindent \textbf{Quantum 3WM Hamiltonian}
We analyze the 3WM process using the canonical quantization method \cite{Hillery-Mlodinow-canonical-quantization-nonlinear-medium, Drummond-dispersive-nonlinear-quantization, Hillery-quantum-nonlinear-optics-rev}. For simplicity, the electric field polarisation vectors are assumed to lie along the 3WM crystal axes, as for a periodically poled or otherwise noncritically phase-matched crystal. We assume conservation of momentum and energy for the carrier waves and retain only phase-matched processes. The $\chi^{(2)}$ nonlinear Hamiltonian is then
\begin{equation}
H_\mathrm{3WM} = i \hbar \alpha \int dz \: \Psi_1^\dagger \Psi_2 \Psi_3 + \mbox{h.c.} \label{nlham}
\end{equation}
\noindent where the coupling constant $\alpha$ is determined by the material nonlinear susceptibility and the beam geometry and is readily calculated in the classical limit. The escort field is taken to be a classical field of constant intensity that is phase-modulated to impart the desired spectral modulation to mode 2. We write $\Psi_3 = \xi  \exp [i \phi(z + v_e t)]$, where $\xi$ is the (real, positive) amplitude of the classical escort field and $\phi(z + v_e t)$ is the local phase. With the definition $\Omega \equiv \alpha \xi$, Eq. (\ref{nlham}) reduces to Eq. (\ref{hamid})
for the case $v_e = 0$. \\

\noindent \textbf{Phase functions for waveform shaping}

For simplicity, we assume that the input mode and the desired target mode are both transform-limited. To match the power spectrum of the 3WM product to the desired spectrum, $\phi(z)$ should satisfy $|\tilde{\alpha}_2(k)| = \left| \int dz \: \alpha_1(z) e^{i \phi(z)} e^{-i k z} \right|$, where $\tilde{\alpha}_2(k)$ is the Fourier transform of $\alpha_2(z)$. In general, one can satisfy this constraint by numerical least-squares minimization. However, when the input and target bandwidths differ substantially, the method of stationary phase applies to the integral and $\phi(z)$ has a closed-form solution in this limit when the target is Gaussian. Writing $\tilde{\alpha}_2(k) \propto e^{-k^2/(2 \sigma^2)}$, we find
\begin{equation}
\phi(z) \approx \frac{\sqrt{2}}{\sigma} \int_{-\infty}^z d\zeta_1 \: \erf^{-1} \left[ a + b \int_0^{\zeta_1} \!\! d\zeta_2 \: \alpha_1^2(\zeta_2) \right] \label{phieq} \\
\end{equation}
\noindent where $\erf$ is the error function, $f^{-1}(x) = y$ denotes the solution of $f(y) = x$, and the constants $a, b$ are set by the boundary conditions of the transformation. After 3WM, the phase of $\tilde{\alpha}_2(k)$ is nontrivial and the 3WM product pulse is therefore not transform-limited. The output pulse shaper applies a spectral compensation phase $\gamma(k)$, implementing the unitary transformation
\begin{equation} \label{Psiout}
\Psi_\mathrm{out}(z) = \frac{1}{2\pi} \int dz \: \Psi_2(z, t=T) \int dk \: e^{i \gamma(k)} e^{i k (\zeta-z)}
\end{equation}
Choosing
\begin{equation}
\gamma(k) = -\phi\left( \alpha_1^{-1} \left( \frac{1}{\sqrt{b}} \: e^{-k^2/\sigma^2} \right) \right) \label{gameq}
\end{equation}
removes the unwanted phase, so that the output pulse is transform-limited with the desired spectrum. When $\alpha_1(z) \propto e^{-z^2/(2 \mu^2)}$ is also Gaussian, equations (\ref{phieq}) and (\ref{gameq}) reduce to $\phi(z) = \sigma z^2/ (2 \mu)$ and $\gamma(k) = -\mu k^2/(2 \sigma)$, while for a single-sided exponential the solution is that given in Eq. (\ref{Gaustarg}). \\

\noindent \textbf{Dispersive evolution and error in state transfer}

Errors in the state transfer arise from group-velocity mismatch between the three modes in the 3WM medium. In the comoving frame with velocity $\bar{v} = (v_1+v_2)/2$, the quantum Hamiltonian for group-velocity mismatch (GVM) and group-velocity dispersion (GVD) can be written as \cite{Drummond-Corney-quantum-noise-fibers}
\begin{equation}
H_\mathrm{disp} = \sum_{j=1,2} \int dz \: \left[ \frac{i \hbar v_j}{2} \frac{\partial \Psi_j^\dagger}{\partial z} \Psi_j + \frac{\beta_j}{4} \frac{\partial \Psi_j^\dagger}{\partial z} \frac{\partial \Psi_j}{\partial z} \right] + \mbox{h.c.}
\end{equation}
while the 3WM Hamiltonian (\ref{hamid}) is also modified because the phase $\phi(z)$ of the escort field phase evolves under dispersion (see Supplementary Discussion). Moving to the interaction picture with respect to the original 3WM Hamiltonian (\ref{hamid}), one derives the
additional unitary evolution due to dispersive effects, $U_\mathrm{disp}$. A second-order Dyson series solution for $U_\mathrm{disp}$ shows that GVM mixes the vacuum noise of the initially unoccupied mode 2 into the state transfer, while GVD has a negligible effect. The removal of phase by the output pulse shaper just implements a unitary transformation on the 3WM output field, which has no effect on the fidelity. The fidelity is then evaluated as $F = \big| \langle U_\mathrm{disp} \rangle \big|$, where the expectation value is taken with respect to the initial states of modes 1 and 2 and any other systems entangled with mode 1. Compensating the phase according to Eq. (\ref{phscomp}) is found to minimise the error independently of the input state. Taking an initial pure state in mode 1 and the vacuum state in mode 2, we obtain Eq. (\ref{fidelity}). \\

\noindent \textbf{Numerical simulations of error in state transfer}

The Heisenberg equations of motion for the field operators $\Psi_1$, $\Psi_2$ are linear, so the operator of the target field after time $t$ will be a linear combination of the initial field operators.  If the input pulse has a single spatial mode $A(z)$, such that $\left|\psi(0)\right> =  f[a^\dagger] \left| 0 \right> =  f\left[\int dx A(z) \Psi_1^\dagger(z)\right] \left| 0 \right> = \sum_n c_n \ket{n}$, the quantum state at time $t$ is
\begin{equation}
\left|\psi(0)\right> =  f\left[ \int dx \left\{ A_1(z,t) \Psi_1^\dagger(x) + A_2(z,t) \Psi_2^\dagger(x)\right\}\right] \left| 0 \right>,
\end{equation}
with $A_1(z,0) = A(z)$ and $A_2(z,0) = 0$. The $A_n$ obey the same linear equations as $\Psi_n$, but are $c$-number amplitudes rather than operators. Simulating these equations allows us to calculate the fidelity as
\begin{equation}
F = \left| \sum_n |c_n|^2  \left [ \int dz A^*(z)e^{-i\phi(z)} A_2(z,T) \right ]^n \right|
\end{equation}
For fidelities close to unity and a correctly compensated phase, this reproduces the linear dependence on $\langle n \rangle$ found in the perturbative calculation (\ref{fidelity}). For definiteness, we only show results for a single-photon input state.

\end{small}

\acknowledgments

This work was supported by the Australian Research Council under DP0773354 (Kielpinski), CE0348250 (Wiseman), FF0458313 (Wiseman), and CE0348178 (Corney). We thank Geoff Pryde for helpful conversations. \\

\noindent \textbf{Author contributions}

\noindent D.K. originated the scheme and wrote most of the manuscript. H.M.W. and D.K. calculated perturbative results for the error. J.F.C. and D.K. performed the numerical simulations. \\

\newpage

\begin{center}
\textbf{Supplementary discussion: Perturbative calculation of waveform conversion error}
\end{center}

\section*{Dispersive evolution}

Error in the state transfer arises from dispersion in the 3WM evolution. To calculate the error, we treat all dispersion-related terms as perturbations to the nondispersive 3WM Hamiltonian
\begin{equation}
H_\mathrm{3WM} \equiv i \hbar \Omega \int dz \Psi_1^\dagger(z,t) \Psi_2(z,t) e^{i\phi(z)} + \mbox{h.c.}
\end{equation}
which induces ideal state transfer by the unitary evolution $U_0(T)$ over the time $T = \pi/(2 \Omega)$. Mode 2 is assumed to be initially unoccupied throughout the calculation, as for the applications discussed in the main paper. \\

We consider dispersion up to second order, i.e., group velocity mismatch (GVM) between modes 1, 2, and 3, and group velocity dispersion (GVD) within each of these modes. In nondegenerate mixing, as considered here, GVM has much larger effects than GVD. The GVD and GVM parameters are expressed using the variables $v = (v_1 - v_2)/2$ and $v_e = v_3-(v_1+v_2)/2$ for GVM (where $v_i = d\omega_i/dk_i$) and $\beta_i = d^2\omega_i/dk_i^2$ for GVD. We define $\epsilon \ll 1$ the perturbative expansion parameter and take $v, v_e \sim {\cal O}(\epsilon)$, $\beta_i \sim {\cal O}(\epsilon^2)$. The dispersion of the quantum modes 1 and 2 is described by the Hamiltonian
\begin{align}
H_\mathrm{disp} &\equiv \frac{i \hbar v}{2} \int dz  \left[\frac{\partial \Psi_1^\dagger(z,t)}{\partial z} \Psi_1(z,t) - \frac{\partial \Psi_2^\dagger(z,t)}{\partial z} \Psi_2(z,t) + \mbox{h.c.} \right] \\
&\hspace{3cm} + \frac{\hbar}{4} \int dz \left[ \beta_1 \frac{\partial^2 \Psi_1^\dagger(z,t)}{\partial z^2} \Psi_1(z,t) + \beta_2 \frac{\partial^2 \Psi_2^\dagger(z,t)}{\partial z^2} \Psi_2(z,t) + \mbox{h.c.} \right]
\end{align}
The dispersion of the escort laser (mode 3) affects the quantum state transfer indirectly through the 3WM Hamiltonian. At ${\cal O}(\epsilon^2)$, the escort mode is governed by the evolution equation
\begin{equation}
\frac{d\Psi_3}{dt} = - v_e \frac{d\Psi_3}{dz} + \frac{i \beta_3}{2} \frac{d^2\Psi_3}{dz^2} \label{escdisp}
\end{equation}

As will be seen, the error can be minimised by precompensation of the phase function. For an escort pulse of constant intensity, $\Psi_3 \propto e^{-i \chi(z,t)}$, we write the compensated phase function at $t=0$ as $\chi(z,0) = \phi(z) + \Delta(z)$. The solution of Eq. (\ref{escdisp}) is
\begin{equation}
\chi(z,t) = \phi(z - v_e t) + \Delta(z - v_e t) - \frac{\beta_3 t}{2} \left[ (\phi'(z))^2 - i \phi''(z) \right]
\end{equation}
where the prime applied to functions (as in $\phi'$) indicates differentiation with respect to $z$. Including the effect of dispersion, the 3WM Hamiltonian becomes
\begin{align}
H^{(d)}_\mathrm{3WM} &\equiv H_\mathrm{3WM} + \left[ i \hbar \Omega \int dz \Psi_1^\dagger(z,t) \Psi_2(z,t) \left( e^{i\chi(z,t)} - e^{i \phi(z)} \right) + \mbox{h.c.} \right]
\end{align}

We go to an interaction frame with respect to $H_\mathrm{3WM}$ to obtain the perturbation Hamiltonian
\begin{equation}
V_I(t) \equiv U_0^\dagger (H_\mathrm{disp} + H^{(d)}_\mathrm{3WM} - H_\mathrm{3WM}) U_0 \approx V_1(t) + V_2(t)
\end{equation}
where $V_1(t) = {\cal O}(\epsilon)$ and $V_2(t) = {\cal O}(\epsilon^2)$. The Dyson series gives the unitary operator $U_I(t)$ describing the perturbed evolution in the interaction frame:
\begin{equation}
U_I(t) = 1 - \frac{i}{\hbar} \int_0^t dt \: V_I(t) - \frac{1}{\hbar^2} \int_0^t dt' \: \int_0^{t'} dt'' \: V_I(t') V_I(t'') + \ldots \label{dyson}
\end{equation}
Expanding the Dyson series to $\mathcal{O}(\epsilon^2)$ gives
\begin{gather}
U_\mathrm{eff}(t) \approx 1 + U_{11}(t) + U_{21}(t) + U_{12}(t) \\
U_{11}(t) \equiv - \frac{i}{\hbar} \int_0^t dt \: V_1(t) \qquad U_{21}(t) \equiv  - \frac{i}{\hbar} \int_0^t dt \: V_2(t) \qquad U_{12}(t) \equiv  - \frac{1}{\hbar^2} \int_0^t dt' \int_0^{t'} dt'' \: V_1(t') V_1(t'')
\end{gather}
It can be shown that $V_1$ and $V_2$ are both Hermitian, so the expectation values of $U_{11}(T)$ and $U_{21}(T)$ are purely imaginary. $U_{11}(T)$ contributes to the error only as ${\cal O}(v^2,v^2_e) \sim {\cal O}(\epsilon^2)$, showing that we must include $U_{12}(T)$ in a consistent perturbative expansion of the error to lowest order in GVM. Moreover, $U_{21}(T)$ contributes only as ${\cal O}(\beta^2_i) \sim {\cal O}(\epsilon^4)$. We consider only error terms up to ${\cal O}(\epsilon^2)$, so we discard $U_{21}(T)$. Hence GVD has no effect on the error in this analysis. \\

The solution of the nondispersive Hamiltonian lets us express $\Psi_1(z,t)$ and $\Psi_2(z,t)$ in terms of $\Phi(z) \equiv \Psi_1(z,t=0)$ and $\Upsilon(z) \equiv \Psi_2(z,t=0)$, giving
\begin{align}
V_1(t) &= - \frac{\hbar}{2} \int dz \left[ \left( \Omega \sin 2 \Omega t \: \Delta + \left( v \sin^2 \Omega t - v_e \Omega t \sin 2 \Omega t \right) \phi' \right) \Phi^\dagger \Phi - i v \cos 2 \Omega t \: \frac{d\Phi^\dagger}{dz} \Phi \right. \nonumber \\
& \hspace{3cm} \left. - 2 \Omega e^{i \phi} \cos 2 \Omega t \left( \Delta - v_e t \phi' \right) \Phi^\dagger \Upsilon + i e^{i \phi} v \sin 2 \Omega t \left( \frac{d \Phi^\dagger}{dz} \Upsilon - \Phi^\dagger \frac{d \Upsilon}{d z} \right) \right] + \mbox{h.c.}
\end{align}
Since $V_1(t)$ is normally ordered and mode 2 is initially unoccupied, terms that involve only mode 2 do not affect the fidelity and are omitted from the expressions. We wish to evaluate the error at the end of the state transfer, which occurs at time $t = T \equiv \pi/(2\Omega)$, giving
\begin{align}
U_{11}(T) &= -\frac{i}{4 \Omega} \int dz \: h(z) \Phi^\dagger(z) \Phi(z) \\
h(z) &\equiv 4 \Omega \Delta(z) + \pi (v - v_e) \phi'(z)
\end{align}
so that $\Delta \sim {\cal O}(\epsilon)$ or higher if $\Delta$ is to minimise the error in state transfer. The remaining term in $U_\mathrm{eff}(T)$ is computed to be
\begin{align}
U_{12}(T) &= \frac{1}{64 \Omega^2} \int dz_1 dz_2 \left\{ -2 h(z_1) h(z_2) \Phi^\dagger(z_1) \Phi(z_1) \Phi^\dagger(z_2) \Phi(z_2) \right. \nonumber \\
&\hspace{2cm} \left. - 8 v_e^2 \cos\left[\phi(z_1)-\phi(z_2)\right] \phi'(z_1) \phi'(z_2) \Phi^\dagger(z_1) \Upsilon(z_1) \Upsilon^\dagger(z_2) \Phi(z_2) \right. \nonumber \\
&\hspace{2cm} \left. - i v e^{i(\phi(z_1)-\phi(z_2))} \left[ g_+(z_1) \Phi^\dagger(z_1) \Upsilon(z_1) \left( \frac{d\Upsilon^\dagger(z_2)}{dz_2} \Phi(z_2) - \Upsilon^\dagger(z_2) \frac{d\Phi(z_2)}{dz_2} \right) \right. \right. \nonumber \\
&\hspace{6cm} \left. \left. - g_-(z_2) \left( \frac{d\Phi^\dagger(z_1)}{dz_1} \Upsilon(z_1) - \Phi^\dagger(z_1) \frac{d\Upsilon(z_1)}{dz_1} \right) \Upsilon^\dagger(z_2) \Phi(z_2) \right] \right. \nonumber \\
&\hspace{2cm} \left. - 8 v^2 e^{i(\phi(z_1)-\phi(z_2))} \left[ \frac{d\Phi^\dagger(z_1)}{dz_1} \Upsilon(z_1) \Upsilon^\dagger(z_2) \frac{d\Phi(z_2)}{dz_2} + \Phi^\dagger(z_1) \frac{d\Upsilon(z_1)}{dz_1} \frac{d\Upsilon^\dagger(z_2)}{dz_2} \Phi(z_2) \right. \right. \nonumber \\
&\hspace{6cm} \left. \left. - \Phi^\dagger(z_1) \frac{d\Upsilon(z_1)}{dz_1} \Upsilon^\dagger(z_2) \frac{d\Phi(z_2)}{dz_2} - \frac{d\Phi^\dagger(z_1)}{dz_1} \Upsilon(z_1) \frac{d\Upsilon^\dagger(z_2)}{dz_2} \Phi(z_2) \right] \right\} \label{eq:u12}
\end{align}
Here $g_\pm(z) \equiv 4 \pi \Omega \Delta(z) - (\pi^2 \pm 8) v_e \phi'(z)$ and we have eliminated terms with purely imaginary expectation values, as these terms do not contribute to the error at ${\cal O}(\epsilon^2)$.

\section*{Fidelity calculation}

We quantify the error in the state transfer by computing the fidelity $F$ between the actual output state of the waveform converter and the ideal dispersion-free output state. The final pulse shaping just implements a unitary transformation on the 3WM output field, which has no effect on the fidelity. For many applications one wishes to convert entangled states involving both mode 1 and some other modes, so the full state before waveform conversion takes the form
\begin{equation}
\ket{\Psi_i} = \sum_j \alpha_j \ket{\psi_i^{(j)}} \otimes \ket{\chi^{(j)}}
\end{equation}
where the $\ket{\psi_i^{(j)}}$ are orthonormal states of mode 1 and the $\ket{\chi^{(j)}}$ are orthonormal states over the other systems. Writing $U_\mathrm{full}(T)$ as the unitary evolution under the full dispersive Hamiltonian, we have
\begin{align}
F &= \sum_j |\alpha_j|^2 \left| \Braket{\psi_i^{(j)} | U_\mathrm{full}^\dagger U_0 | \psi_i^{(j)}} \right|
\end{align}
Thus, since we can always find the transfer fidelity for entangled input states by taking an appropriate weighted sum over pure-state fidelities, we need only calculate the transfer fidelity of a pure input state $\ket{\psi_i}$. In many cases, such a state is characterised by a mode creation operator
\begin{equation}
a^\dagger = \int dz\: A(z) \Phi^\dagger(z)
\end{equation}
where $\Phi(z) = \Psi_1(z, t=0)$ and the mode wavefunction $A(z)$ is normalised as $\int dz\: |A(z)|^2 = 1$. An initial $k$-photon number state in the mode is given by $(a^\dagger)^k \ket{0}$ and a general single-mode initial pure state $\ket{\psi_i} = \sum_{k=0}^\infty c_k \ket{k}$ can be written as $\ket{\psi_i} = f(a^\dagger) \ket{0}$, where $\ket{k}$ denotes a number state and $f(x)$ is defined through the series expansion $f(x) = \sum_{k=0}^\infty c_k x^k/\sqrt{k!}$. \\

The pure-state fidelity is given by the squared overlap between the state obtained from phase-compensated dispersive evolution and that obtained from  uncompensated nondispersive evolution. In an interaction frame with respect to the nondispersive 3WM Hamiltonian, we have
\begin{align}
F &= \left| \Braket{\psi_i | U_\mathrm{eff}^\dagger(T) | \psi_i} \right|
\end{align}
and using the matrix elements derived in the Appendix, we find
\begin{align}
\Braket{\psi_i | U_{11}(T) | \psi_i} &= \frac{i \langle n \rangle}{4 \Omega} \int dz \: h(z) |A(z)|^2 \\
\Braket{\psi_i | U_{12}(T) | \psi_i} &= \frac{1}{64 \Omega^2} \int dz_1 dz_2 \left\{ -2 \langle n(n-1) \rangle h(z_1) h(z_2) |A(z_1)|^2 |A(z_2)|^2 + 2 \langle n \rangle h(z_1) h(z_2) A^*(z_1) A(z_2) \delta(z_1-z_2) \right. \nonumber \\
&\hspace{1cm} \left. - 8 \langle n \rangle v_e^2 \cos\left[\phi(z_1)-\phi(z_2)\right] \phi'(z_1) \phi'(z_2) A^*(z_1) A(z_2) \delta(z_1-z_2) \right. \nonumber \\
&\hspace{1cm} \left. - i \langle n \rangle v e^{i(\phi(z_1)-\phi(z_2))} \left[ g_+(z_1) \left( A^*(z_1) A(z_2) \partial_{z_2} \delta(z_1-z_2) - A^*(z_1) A'(z_2) \delta(z_1-z_2) \right) \right. \right. \nonumber \\
&\hspace{5cm} \left. \left. - g_-(z_2) \left( A^{\prime *}(z_1) A(z_2) \delta(z_1-z_2) - A^*(z_1) A(z_2) \partial_{z_1} \delta(z_1-z_2) \right) \right] \right. \nonumber \\
&\hspace{1cm} \left. - 8 \langle n \rangle v^2 e^{i(\phi(z_1)-\phi(z_2))} \left[ A^{\prime *}(z_1) A'(z_2) \delta(z_1-z_2) + A^*(z_1) A(z_2) \partial_{z_1} \partial_{z_2} \delta(z_1-z_2) \right. \right. \nonumber \\
&\hspace{5cm} \left. \left. - A^*(z_1) A'(z_2) \partial_{z_1} \delta(z_1-z_2) - A^{\prime *}(z_1) A(z_2) \partial_{z_2} \delta(z_1-z_2) \right] \right\} \\
&= \frac{1}{64 \Omega^2} \left\{ -2 \langle n(n-1) \rangle \left( \int dz \: h |A|^2 \right)^2 + 2 \langle n \rangle \int dz \: h^2 |A|^2 - 8 \langle n \rangle v_e^2 \int dz |\phi' A|^2 \right. \nonumber \\
&\hspace{2cm} \left. + 8 \langle n \rangle v v_e \int dz \: \phi' \left[ A^* (i A' + \phi' A) + \mbox{h.c.} \right] - 8 \langle n \rangle v^2 \int dz \: |2 A' - i \phi' A|^2 \right\} \\
&= \frac{1}{32 \Omega^2} \left\{ - \langle n(n-1) \rangle \left( \int dz \: h |A|^2 \right)^2 + \langle n \rangle \int dz \: h^2 |A|^2 - 4 \langle n \rangle \int dz \: \big|2 v A' - i (v + v_e) \phi' A \big|^2 \right\}
\end{align}
where $n$ is the number operator of the mode, $\langle n \rangle$ denotes the expectation value of $n$, and we have retained only real terms. The fidelity is optimised when $h(z) = 0$ so that
\begin{align}
\Delta_\mathrm{opt}(z) &= \frac{\pi (v_e - v) \phi'(z)}{4 \Omega} \\
F_\mathrm{opt} &= 1 - \frac{\langle n \rangle}{8 \Omega^2} \int dz \: \big| 2 v A'(z) - i (v + v_e) \phi'(z) A(z) \big|^2
\end{align}
which is just Eq. (\ref{fidelity}).

\section*{Appendix: Matrix elements for pure states}

To evaluate the matrix elements involved in the fidelity calculation, we first observe that
\begin{align}
[\Phi(z),a^\dagger] &= \int d\zeta \: A(\zeta) \: [\Phi(z), \Phi^\dagger(\zeta)] = \int d\zeta \: A(\zeta) \delta(\zeta-z) = A(z) \\
[a, a^\dagger] &= \int d\zeta \: A^*(\zeta) \: [\Phi(\zeta), a^\dagger] = \int d\zeta \: |A(\zeta)|^2 = 1 \\
[\Phi(z),(a^\dagger)^k] &= k A(z) (a^\dagger)^{k-1} \\
\Braket{0 | a^j (a^\dagger)^k | 0} &= k! \delta_{jk} \label{eq:apwrs}
\end{align}

Writing $\langle n \rangle$ for the expectation value of the photon number in $\ket{\psi_i}$, we find
\begin{align}
{\cal M}_1 \equiv \Braket{\psi_i | \Phi^\dagger(z_1) \Phi(z_2) | \psi_i} &= \sum_{k=0}^\infty \sum_{j=0}^\infty \frac{c_k^* c_j}{\sqrt{k!j!}} \Braket{0 | a^{j} \Phi^\dagger(z_1) \Phi(z_2) (a^\dagger)^{k} | 0} \\
&= A^*(z_2) A(z_1) \sum_{k=0}^\infty \frac{|c_k|^2}{k!} k^2 \Braket{0 | a^{k-1} (a^\dagger)^{k-1} | 0}\\
&= A^*(z_1) A(z_2) \sum_{k=0}^\infty k |c_k|^2 \\
&= \langle n \rangle A^*(z_1) A(z_2) \\
{\cal M}_2 \equiv \Braket{\psi_i | \Phi^\dagger(z_1) \Phi(z_2) \Phi^\dagger(z_3) \Phi(z_4) | \psi_i} &= \sum_{k=0}^\infty \sum_{j=0}^\infty \frac{c_k^* c_j}{\sqrt{k!j!}} \Braket{0 | a^{j} \Phi^\dagger(z_1) \Phi(z_2) \Phi^\dagger(z_3) \Phi(z_4) (a^\dagger)^{k} | 0} \\
&\hspace{-2cm}= A^*(z_1) A(z_4) \sum_{k=0}^\infty \sum_{j=0}^\infty \frac{c_k^* c_j}{\sqrt{k!j!}} j k \Braket{0 | a^{j-1} \Phi(z_2) \Phi^\dagger(z_3) (a^\dagger)^{k-1} | 0} \\
&\hspace{-2cm}= A^*(z_1) A(z_4) \sum_{k=0}^\infty \sum_{j=0}^\infty \frac{c_k^* c_j}{\sqrt{k!j!}} j k \Braket{0 | a^{j-1} \left( \Phi^\dagger(z_3) \Phi(z_2) + \delta(z_2-z_3) \right) (a^\dagger)^{k-1} | 0} \\
&\hspace{-2cm}= A^*(z_1) A(z_4) \sum_{k=0}^\infty \frac{|c_k|^2}{k!} k^2 \left( (k-1)^2 (k-2)! A^*(z_3) A(z_2) + \delta(z_2-z_3) (k-1)! \right) \\
&\hspace{-2cm}= \langle n(n-1) \rangle A^*(z_1) A(z_2) A^*(z_3) A(z_4) + \langle n \rangle A^*(z_1) A(z_4) \delta(z_2-z_3)
\end{align}
Then
\begin{align}
\Braket{\psi_i | \Phi^\dagger(z_1) \frac{d\Phi(z_2)}{dz_2} | \psi_i} &= \partial_{z_1} {\cal M}_1 = \langle n \rangle A^*(z_1) A'(z_2) \\
\Braket{\psi_i | \frac{d\Phi^\dagger(z_1)}{dz_!} \frac{d\Phi(z_2)}{dz_2} | \psi_i} &= \partial_{z_1} \partial_{z_2} {\cal M}_1 = \langle n \rangle (A^*)'(z_1) A'(z_2) \\
\Braket{\psi_i | \Phi^\dagger(z_1) \Phi(z_2) \Phi^\dagger(z_3) \frac{d\Phi(z_4)}{dz_4} | \psi_i} &= \partial_{z_4} {\cal M}_2 = \langle n(n-1) \rangle A^*(z_1) A(z_2) A^*(z_3) A'(z_4) - \langle n \rangle A^*(z_1) A'(z_4) \delta(z_2-z_3)
\end{align}
and similarly for the other matrix elements involving mode 1. For mode 2, which is initially unoccupied, we calculate
\begin{align}
{\cal M}_3 \equiv \Braket{0 | \Upsilon(z_1) \Upsilon^\dagger(z_2) | 0} &= \delta(z_1-z_2) \\
\Braket{0 | \frac{d\Upsilon(z_1)}{dz_1} \Upsilon^\dagger(z_2) | 0} &= \partial_{z_1} {\cal M}_3 = \partial_{z_1} \delta(z_1-z_2) \\
\Braket{0 | \frac{d\Upsilon(z_1)}{dz_1} \frac{d\Upsilon^\dagger(z_2)}{dz_2} | 0} &= \partial_{z_1} \partial_{z_2} \delta(z-z_2)
\end{align}


\begin{thebibliography}{29}
\expandafter\ifx\csname natexlab\endcsname\relax\def\natexlab#1{#1}\fi
\expandafter\ifx\csname bibnamefont\endcsname\relax
  \def\bibnamefont#1{#1}\fi
\expandafter\ifx\csname bibfnamefont\endcsname\relax
  \def\bibfnamefont#1{#1}\fi
\expandafter\ifx\csname citenamefont\endcsname\relax
  \def\citenamefont#1{#1}\fi
\expandafter\ifx\csname url\endcsname\relax
  \def\url#1{\texttt{#1}}\fi
\expandafter\ifx\csname urlprefix\endcsname\relax\def\urlprefix{URL }\fi
\providecommand{\bibinfo}[2]{#2}
\providecommand{\eprint}[2][]{\url{#2}}

\bibitem[{\citenamefont{Briegel et~al.}(1998)\citenamefont{Briegel, Dur, Cirac,
  and Zoller}}]{Briegel-Zoller-quantum-repeater}
\bibinfo{author}{\bibfnamefont{H.-J.} \bibnamefont{Briegel}},
  \bibinfo{author}{\bibfnamefont{W.}~\bibnamefont{Dur}},
  \bibinfo{author}{\bibfnamefont{J.~I.} \bibnamefont{Cirac}}, \bibnamefont{and}
  \bibinfo{author}{\bibfnamefont{P.}~\bibnamefont{Zoller}},
  \bibinfo{journal}{Phys. Rev. Lett.} \textbf{\bibinfo{volume}{81}},
  \bibinfo{pages}{5932} (\bibinfo{year}{1998}).

\bibitem[{\citenamefont{Duan et~al.}(2001)\citenamefont{Duan, Lukin, Cirac, and
  Zoller}}]{Duan-Zoller-linear-optics-atom-QC}
\bibinfo{author}{\bibfnamefont{L.-M.} \bibnamefont{Duan}},
  \bibinfo{author}{\bibfnamefont{M.~D.} \bibnamefont{Lukin}},
  \bibinfo{author}{\bibfnamefont{J.~I.} \bibnamefont{Cirac}}, \bibnamefont{and}
  \bibinfo{author}{\bibfnamefont{P.}~\bibnamefont{Zoller}},
  \bibinfo{journal}{Nature} \textbf{\bibinfo{volume}{414}},
  \bibinfo{pages}{413} (\bibinfo{year}{2001}).

\bibitem[{\citenamefont{Gisin et~al.}(2002)\citenamefont{Gisin, Ribordy,
  Tittel, and Zbinden}}]{Gisin-Zbinden-quantum-crypto-rev}
\bibinfo{author}{\bibfnamefont{N.}~\bibnamefont{Gisin}},
  \bibinfo{author}{\bibfnamefont{G.}~\bibnamefont{Ribordy}},
  \bibinfo{author}{\bibfnamefont{W.}~\bibnamefont{Tittel}}, \bibnamefont{and}
  \bibinfo{author}{\bibfnamefont{H.}~\bibnamefont{Zbinden}},
  \bibinfo{journal}{Rev. Mod. Phys.} \textbf{\bibinfo{volume}{74}},
  \bibinfo{pages}{145} (\bibinfo{year}{2002}).

\bibitem[{\citenamefont{Wootters and Zurek}(1982)}]{Wootters-Zurek-no-cloning}
\bibinfo{author}{\bibfnamefont{W.~K.} \bibnamefont{Wootters}} \bibnamefont{and}
  \bibinfo{author}{\bibfnamefont{W.~H.} \bibnamefont{Zurek}},
  \bibinfo{journal}{Nature} \textbf{\bibinfo{volume}{299}},
  \bibinfo{pages}{802} (\bibinfo{year}{1982}).

\bibitem[{\citenamefont{Blinov et~al.}(2004)\citenamefont{Blinov, Moehring,
  Duan, and Monroe}}]{Blinov-Monroe-ion-photon-entanglement}
\bibinfo{author}{\bibfnamefont{B.~B.} \bibnamefont{Blinov}},
  \bibinfo{author}{\bibfnamefont{D.~L.} \bibnamefont{Moehring}},
  \bibinfo{author}{\bibfnamefont{L.-M.} \bibnamefont{Duan}}, \bibnamefont{and}
  \bibinfo{author}{\bibfnamefont{C.}~\bibnamefont{Monroe}},
  \bibinfo{journal}{Nature} \textbf{\bibinfo{volume}{428}},
  \bibinfo{pages}{153} (\bibinfo{year}{2004}).

\bibitem[{\citenamefont{McKeever et~al.}(2004)\citenamefont{McKeever, Boca,
  Boozer, Miller, Buck, Kuzmich, and
  Kimble}}]{McKeever-Kimble-CQED-single-photon}
\bibinfo{author}{\bibfnamefont{J.}~\bibnamefont{McKeever}},
  \bibinfo{author}{\bibfnamefont{A.}~\bibnamefont{Boca}},
  \bibinfo{author}{\bibfnamefont{A.~D.} \bibnamefont{Boozer}},
  \bibinfo{author}{\bibfnamefont{R.}~\bibnamefont{Miller}},
  \bibinfo{author}{\bibfnamefont{J.~R.} \bibnamefont{Buck}},
  \bibinfo{author}{\bibfnamefont{A.}~\bibnamefont{Kuzmich}}, \bibnamefont{and}
  \bibinfo{author}{\bibfnamefont{H.~J.} \bibnamefont{Kimble}},
  \bibinfo{journal}{Science} \textbf{\bibinfo{volume}{303}},
  \bibinfo{pages}{1992} (\bibinfo{year}{2004}).

\bibitem[{\citenamefont{Keller et~al.}(2004)\citenamefont{Keller, Lange,
  Hayasaka, Lange, and Walther}}]{Keller-Walther-ion-cavity-single-photon}
\bibinfo{author}{\bibfnamefont{M.}~\bibnamefont{Keller}},
  \bibinfo{author}{\bibfnamefont{B.}~\bibnamefont{Lange}},
  \bibinfo{author}{\bibfnamefont{K.}~\bibnamefont{Hayasaka}},
  \bibinfo{author}{\bibfnamefont{W.}~\bibnamefont{Lange}}, \bibnamefont{and}
  \bibinfo{author}{\bibfnamefont{H.}~\bibnamefont{Walther}},
  \bibinfo{journal}{Nature} \textbf{\bibinfo{volume}{431}},
  \bibinfo{pages}{1075} (\bibinfo{year}{2004}).

\bibitem[{\citenamefont{Chou et~al.}(2004)\citenamefont{Chou, Polyakov,
  Kuzmich, and Kimble}}]{Chou-Kimble-atomic-ensemble-single-photon}
\bibinfo{author}{\bibfnamefont{C.~W.} \bibnamefont{Chou}},
  \bibinfo{author}{\bibfnamefont{S.~V.} \bibnamefont{Polyakov}},
  \bibinfo{author}{\bibfnamefont{A.}~\bibnamefont{Kuzmich}}, \bibnamefont{and}
  \bibinfo{author}{\bibfnamefont{H.~J.} \bibnamefont{Kimble}},
  \bibinfo{journal}{Phys. Rev. Lett.} \textbf{\bibinfo{volume}{92}},
  \bibinfo{pages}{213601} (\bibinfo{year}{2004}).

\bibitem[{\citenamefont{Julsgaard et~al.}(2004)\citenamefont{Julsgaard,
  Sherson, Cirac, Fiur{\'a}{\v{s}}ek, and
  Polzik}}]{Julsgaard-Polzik-atom-quantum-memory-light}
\bibinfo{author}{\bibfnamefont{B.}~\bibnamefont{Julsgaard}},
  \bibinfo{author}{\bibfnamefont{J.}~\bibnamefont{Sherson}},
  \bibinfo{author}{\bibfnamefont{J.~I.} \bibnamefont{Cirac}},
  \bibinfo{author}{\bibfnamefont{J.}~\bibnamefont{Fiur{\'a}{\v{s}}ek}},
  \bibnamefont{and} \bibinfo{author}{\bibfnamefont{E.~S.}
  \bibnamefont{Polzik}}, \bibinfo{journal}{Nature}
  \textbf{\bibinfo{volume}{432}}, \bibinfo{pages}{482} (\bibinfo{year}{2004}).

\bibitem[{\citenamefont{Hosseini et~al.}(2009)\citenamefont{Hosseini, Sparkes,
  H{\'e}tet, Longdell, Lam, and
  Buchler}}]{Hosseini-Buchler-coherent-optical-pulse-sequencer}
\bibinfo{author}{\bibfnamefont{M.}~\bibnamefont{Hosseini}},
  \bibinfo{author}{\bibfnamefont{B.~M.} \bibnamefont{Sparkes}},
  \bibinfo{author}{\bibfnamefont{G.}~\bibnamefont{H{\'e}tet}},
  \bibinfo{author}{\bibfnamefont{J.~J.} \bibnamefont{Longdell}},
  \bibinfo{author}{\bibfnamefont{P.~K.} \bibnamefont{Lam}}, \bibnamefont{and}
  \bibinfo{author}{\bibfnamefont{B.~C.} \bibnamefont{Buchler}},
  \bibinfo{journal}{Nature} \textbf{\bibinfo{volume}{461}},
  \bibinfo{pages}{241} (\bibinfo{year}{2009}).

\bibitem[{\citenamefont{de~Riedmatten et~al.}(2008)\citenamefont{de~Riedmatten,
  Afzelius, Staudt, Simon, and
  Gisin}}]{deRiedmatten-Gisin-rare-earth-photon-storage}
\bibinfo{author}{\bibfnamefont{H.}~\bibnamefont{de~Riedmatten}},
  \bibinfo{author}{\bibfnamefont{M.}~\bibnamefont{Afzelius}},
  \bibinfo{author}{\bibfnamefont{M.~U.} \bibnamefont{Staudt}},
  \bibinfo{author}{\bibfnamefont{C.}~\bibnamefont{Simon}}, \bibnamefont{and}
  \bibinfo{author}{\bibfnamefont{N.}~\bibnamefont{Gisin}},
  \bibinfo{journal}{Nature} \textbf{\bibinfo{volume}{456}},
  \bibinfo{pages}{773} (\bibinfo{year}{2008}).

\bibitem[{\citenamefont{Hedges et~al.}(2010)\citenamefont{Hedges, Longdell, Li,
  and Sellars}}]{Hedges-Sellars-efficient-quantum-memory}
\bibinfo{author}{\bibfnamefont{M.~P.} \bibnamefont{Hedges}},
  \bibinfo{author}{\bibfnamefont{J.~J.} \bibnamefont{Longdell}},
  \bibinfo{author}{\bibfnamefont{Y.}~\bibnamefont{Li}}, \bibnamefont{and}
  \bibinfo{author}{\bibfnamefont{M.~J.} \bibnamefont{Sellars}},
  \bibinfo{journal}{Nature} \textbf{\bibinfo{volume}{465}},
  \bibinfo{pages}{1052} (\bibinfo{year}{2010}).

\bibitem[{\citenamefont{Y{\i}lmaz et~al.}(2010)\citenamefont{Y{\i}lmaz,
  Fallahi, and
  Imamo{\u{g}}lu}}]{Yilmaz-Imamoglu-quantum-dot-photon-entanglement}
\bibinfo{author}{\bibfnamefont{S.~T.} \bibnamefont{Y{\i}lmaz}},
  \bibinfo{author}{\bibfnamefont{P.}~\bibnamefont{Fallahi}}, \bibnamefont{and}
  \bibinfo{author}{\bibfnamefont{A.}~\bibnamefont{Imamo{\u{g}}lu}},
  \bibinfo{journal}{Phys. Rev. Lett.} \textbf{\bibinfo{volume}{105}},
  \bibinfo{pages}{033601} (\bibinfo{year}{2010}).

\bibitem[{\citenamefont{Togan et~al.}(2010)\citenamefont{Togan, Chu, Trifonov,
  Jiang, Maze, Childress, Dutt, S{\o}rensen, Hemmer, Zibrov
  et~al.}}]{Togan-Lukin-NV-photon-entanglement}
\bibinfo{author}{\bibfnamefont{E.}~\bibnamefont{Togan}},
  \bibinfo{author}{\bibfnamefont{Y.}~\bibnamefont{Chu}},
  \bibinfo{author}{\bibfnamefont{A.~S.} \bibnamefont{Trifonov}},
  \bibinfo{author}{\bibfnamefont{L.}~\bibnamefont{Jiang}},
  \bibinfo{author}{\bibfnamefont{J.}~\bibnamefont{Maze}},
  \bibinfo{author}{\bibfnamefont{L.}~\bibnamefont{Childress}},
  \bibinfo{author}{\bibfnamefont{M.~V.~G.} \bibnamefont{Dutt}},
  \bibinfo{author}{\bibfnamefont{A.~S.} \bibnamefont{S{\o}rensen}},
  \bibinfo{author}{\bibfnamefont{P.~R.} \bibnamefont{Hemmer}},
  \bibinfo{author}{\bibfnamefont{A.~S.} \bibnamefont{Zibrov}},
  \bibnamefont{et~al.}, \bibinfo{journal}{Nature}
  \textbf{\bibinfo{volume}{466}}, \bibinfo{pages}{730} (\bibinfo{year}{2010}).

\bibitem[{\citenamefont{Reim et~al.}(2010)\citenamefont{Reim, Nunn, Lorenz,
  Sussman, Lee, Langford, Jaksch, and
  Walmsley}}]{Reim-Walmsley-fast-atom-photon-memory}
\bibinfo{author}{\bibfnamefont{K.~F.} \bibnamefont{Reim}},
  \bibinfo{author}{\bibfnamefont{J.}~\bibnamefont{Nunn}},
  \bibinfo{author}{\bibfnamefont{V.~O.} \bibnamefont{Lorenz}},
  \bibinfo{author}{\bibfnamefont{B.~J.} \bibnamefont{Sussman}},
  \bibinfo{author}{\bibfnamefont{K.~C.} \bibnamefont{Lee}},
  \bibinfo{author}{\bibfnamefont{N.~K.} \bibnamefont{Langford}},
  \bibinfo{author}{\bibfnamefont{D.}~\bibnamefont{Jaksch}}, \bibnamefont{and}
  \bibinfo{author}{\bibfnamefont{I.~A.} \bibnamefont{Walmsley}},
  \bibinfo{journal}{Nature Physics} \textbf{\bibinfo{volume}{4}},
  \bibinfo{pages}{218} (\bibinfo{year}{2010}).

\bibitem[{\citenamefont{McCutcheon et~al.}(2009)\citenamefont{McCutcheon,
  Chang, Zhang, Lukin, and
  Loncar}}]{McCutcheon-Loncar-quantum-pulse-cavity-shaping}
\bibinfo{author}{\bibfnamefont{M.~W.} \bibnamefont{McCutcheon}},
  \bibinfo{author}{\bibfnamefont{D.~E.} \bibnamefont{Chang}},
  \bibinfo{author}{\bibfnamefont{Y.}~\bibnamefont{Zhang}},
  \bibinfo{author}{\bibfnamefont{M.~D.} \bibnamefont{Lukin}}, \bibnamefont{and}
  \bibinfo{author}{\bibfnamefont{M.}~\bibnamefont{Loncar}},
  \bibinfo{journal}{Opt. Express} \textbf{\bibinfo{volume}{17}},
  \bibinfo{pages}{22689} (\bibinfo{year}{2009}).

\bibitem[{\citenamefont{Specht et~al.}(2009)\citenamefont{Specht, Bochmann,
  M{\"u}cke, Weber, Figueroa, Moehring, and
  Rempe}}]{Specht-Rempe-single-photon-phase-shaping}
\bibinfo{author}{\bibfnamefont{H.~P.} \bibnamefont{Specht}},
  \bibinfo{author}{\bibfnamefont{J.}~\bibnamefont{Bochmann}},
  \bibinfo{author}{\bibfnamefont{M.}~\bibnamefont{M{\"u}cke}},
  \bibinfo{author}{\bibfnamefont{B.}~\bibnamefont{Weber}},
  \bibinfo{author}{\bibfnamefont{E.}~\bibnamefont{Figueroa}},
  \bibinfo{author}{\bibfnamefont{D.~L.} \bibnamefont{Moehring}},
  \bibnamefont{and} \bibinfo{author}{\bibfnamefont{G.}~\bibnamefont{Rempe}},
  \bibinfo{journal}{Nature Photon.} \textbf{\bibinfo{volume}{3}},
  \bibinfo{pages}{469} (\bibinfo{year}{2009}).

\bibitem[{\citenamefont{Belthangady et~al.}(2010)\citenamefont{Belthangady,
  Chuu, Yu, Yin, Kahn, and
  Harris}}]{Belthangady-Harris-single-photon-spectral-hiding}
\bibinfo{author}{\bibfnamefont{C.}~\bibnamefont{Belthangady}},
  \bibinfo{author}{\bibfnamefont{C.-S.} \bibnamefont{Chuu}},
  \bibinfo{author}{\bibfnamefont{I.~A.} \bibnamefont{Yu}},
  \bibinfo{author}{\bibfnamefont{G.~Y.} \bibnamefont{Yin}},
  \bibinfo{author}{\bibfnamefont{J.~M.} \bibnamefont{Kahn}}, \bibnamefont{and}
  \bibinfo{author}{\bibfnamefont{S.~E.} \bibnamefont{Harris}},
  \bibinfo{journal}{Phys. Rev. Lett.} \textbf{\bibinfo{volume}{104}},
  \bibinfo{pages}{223601} (\bibinfo{year}{2010}).

\bibitem[{\citenamefont{Vandevender and
  Kwiat}(2004)}]{Vandevender-Kwiat-single-photon-upconversion}
\bibinfo{author}{\bibfnamefont{A.~P.} \bibnamefont{Vandevender}}
  \bibnamefont{and} \bibinfo{author}{\bibfnamefont{P.~G.} \bibnamefont{Kwiat}},
  \bibinfo{journal}{J. Mod. Opt.} \textbf{\bibinfo{volume}{51}},
  \bibinfo{pages}{1433} (\bibinfo{year}{2004}).

\bibitem[{\citenamefont{Albota and
  Wong}(2004)}]{Albota-Wong-telecom-upconversion-detection}
\bibinfo{author}{\bibfnamefont{M.~A.} \bibnamefont{Albota}} \bibnamefont{and}
  \bibinfo{author}{\bibfnamefont{F.~N.~C.} \bibnamefont{Wong}},
  \bibinfo{journal}{Opt. Lett.} \textbf{\bibinfo{volume}{29}},
  \bibinfo{pages}{1449} (\bibinfo{year}{2004}).

\bibitem[{\citenamefont{Langrock et~al.}(2005)\citenamefont{Langrock, Diamanti,
  Roussev, Yamamoto, Fejer, and
  Takesue}}]{Langrock-Fejer-telecom-upconversion-detection}
\bibinfo{author}{\bibfnamefont{C.}~\bibnamefont{Langrock}},
  \bibinfo{author}{\bibfnamefont{E.}~\bibnamefont{Diamanti}},
  \bibinfo{author}{\bibfnamefont{R.~V.} \bibnamefont{Roussev}},
  \bibinfo{author}{\bibfnamefont{Y.}~\bibnamefont{Yamamoto}},
  \bibinfo{author}{\bibfnamefont{M.~M.} \bibnamefont{Fejer}}, \bibnamefont{and}
  \bibinfo{author}{\bibfnamefont{H.}~\bibnamefont{Takesue}},
  \bibinfo{journal}{Opt. Lett.} \textbf{\bibinfo{volume}{30}},
  \bibinfo{pages}{1725} (\bibinfo{year}{2005}).

\bibitem[{\citenamefont{Tanzilli et~al.}(2005)\citenamefont{Tanzilli, Tittel,
  Halder, Alibart, Baldi, Gisin, and
  Zbinden}}]{Tanzilli-Zbinden-entanglement-preserving-upconversion}
\bibinfo{author}{\bibfnamefont{S.}~\bibnamefont{Tanzilli}},
  \bibinfo{author}{\bibfnamefont{W.}~\bibnamefont{Tittel}},
  \bibinfo{author}{\bibfnamefont{M.}~\bibnamefont{Halder}},
  \bibinfo{author}{\bibfnamefont{O.}~\bibnamefont{Alibart}},
  \bibinfo{author}{\bibfnamefont{P.}~\bibnamefont{Baldi}},
  \bibinfo{author}{\bibfnamefont{N.}~\bibnamefont{Gisin}}, \bibnamefont{and}
  \bibinfo{author}{\bibfnamefont{H.}~\bibnamefont{Zbinden}},
  \bibinfo{journal}{Nature} \textbf{\bibinfo{volume}{437}},
  \bibinfo{pages}{116} (\bibinfo{year}{2005}).

\bibitem[{\citenamefont{Bennett and
  Kolner}(1999)}]{Bennett-Kolner-3WM-time-lens-100x}
\bibinfo{author}{\bibfnamefont{C.~V.} \bibnamefont{Bennett}} \bibnamefont{and}
  \bibinfo{author}{\bibfnamefont{B.~H.} \bibnamefont{Kolner}},
  \bibinfo{journal}{Opt. Lett.} \textbf{\bibinfo{volume}{24}},
  \bibinfo{pages}{783} (\bibinfo{year}{1999}).

\bibitem[{\citenamefont{Foster et~al.}(2009)\citenamefont{Foster, Salem,
  Okawachi, Turner-Foster, Lipson, and
  Gaeta}}]{Foster-Gaeta-time-lens-waveform-compression}
\bibinfo{author}{\bibfnamefont{M.~A.} \bibnamefont{Foster}},
  \bibinfo{author}{\bibfnamefont{R.}~\bibnamefont{Salem}},
  \bibinfo{author}{\bibfnamefont{Y.}~\bibnamefont{Okawachi}},
  \bibinfo{author}{\bibfnamefont{A.~C.} \bibnamefont{Turner-Foster}},
  \bibinfo{author}{\bibfnamefont{M.}~\bibnamefont{Lipson}}, \bibnamefont{and}
  \bibinfo{author}{\bibfnamefont{A.~L.} \bibnamefont{Gaeta}},
  \bibinfo{journal}{Nature Photon.} \textbf{\bibinfo{volume}{3}},
  \bibinfo{pages}{581} (\bibinfo{year}{2009}).

\bibitem[{\citenamefont{Yu et~al.}(2002)\citenamefont{Yu, Ro, Cha, Kurimura,
  and Taira}}]{Yu-Taira-PPMgLN-GVM-matching}
\bibinfo{author}{\bibfnamefont{N.~E.} \bibnamefont{Yu}},
  \bibinfo{author}{\bibfnamefont{J.~H.} \bibnamefont{Ro}},
  \bibinfo{author}{\bibfnamefont{M.}~\bibnamefont{Cha}},
  \bibinfo{author}{\bibfnamefont{S.}~\bibnamefont{Kurimura}}, \bibnamefont{and}
  \bibinfo{author}{\bibfnamefont{T.}~\bibnamefont{Taira}},
  \bibinfo{journal}{Opt. Lett.} \textbf{\bibinfo{volume}{27}},
  \bibinfo{pages}{1046} (\bibinfo{year}{2002}).

\bibitem[{\citenamefont{Hillery and
  Mlodinow}(1984)}]{Hillery-Mlodinow-canonical-quantization-nonlinear-medium}
\bibinfo{author}{\bibfnamefont{M.}~\bibnamefont{Hillery}} \bibnamefont{and}
  \bibinfo{author}{\bibfnamefont{L.~D.} \bibnamefont{Mlodinow}},
  \bibinfo{journal}{Phys. Rev. A} \textbf{\bibinfo{volume}{30}},
  \bibinfo{pages}{1860} (\bibinfo{year}{1984}).

\bibitem[{\citenamefont{Drummond}(1990)}]{Drummond-dispersive-nonlinear-quanti%
zation}
\bibinfo{author}{\bibfnamefont{P.~D.} \bibnamefont{Drummond}},
  \bibinfo{journal}{Phys. Rev. A} \textbf{\bibinfo{volume}{42}},
  \bibinfo{pages}{6845} (\bibinfo{year}{1990}).

\bibitem[{\citenamefont{Hillery}(2009)}]{Hillery-quantum-nonlinear-optics-rev}
\bibinfo{author}{\bibfnamefont{M.}~\bibnamefont{Hillery}},
  \bibinfo{journal}{Acta Physica Slovaca} \textbf{\bibinfo{volume}{59}},
  \bibinfo{pages}{1} (\bibinfo{year}{2009}).

\bibitem[{\citenamefont{Drummond and
  Corney}(2001)}]{Drummond-Corney-quantum-noise-fibers}
\bibinfo{author}{\bibfnamefont{P.~D.} \bibnamefont{Drummond}} \bibnamefont{and}
  \bibinfo{author}{\bibfnamefont{J.~F.} \bibnamefont{Corney}},
  \bibinfo{journal}{J. Opt. Soc. Am. B} \textbf{\bibinfo{volume}{18}},
  \bibinfo{pages}{139} (\bibinfo{year}{2001}).

\end{thebibliography}
\end{document}